\documentclass[aps,prl,preprint,superscriptaddress]{revtex4}
\usepackage{graphicx}
\usepackage{dcolumn}
\usepackage{bm}
\usepackage{psfig}

\newcommand{\be}{\begin{equation}}
\newcommand{\ee}{\end{equation}}
\newcommand{\bea}{\begin{eqnarray}}
\newcommand{\eea}{\end{eqnarray}}
\newcommand{\bean}{\begin{eqnarray*}}
\newcommand{\eean}{\end{eqnarray*}}

\newcommand {\sla}[1]{ #1 \!\!\!/}

\begin{document}

\title{\boldmath Mass and $K\Lambda$ coupling of $N^*(1535)$}

\author{
B.C.Liu$^{1}$, B.~S.~Zou$^{2,1}$\\
$^1$ Institute of High Energy Physics, P.O.Box 918(4), Beijing 100049, China\\
$^2$ CCAST (World Lab.), P.O.Box 8730, Beijing 100080, China}

\date{March 29, 2005}

\begin{abstract}
Using resonance isobar model and effective Lagrangian approach,
from recent BES results on $J/\psi\to\bar pp\eta$ and $\psi\to\bar
pK^+\Lambda$, we deduce the ratio between effective coupling
constants of $N^*(1535)$ to $K\Lambda$ and $p\eta$  to be $R\equiv
g_{N^*(1535)K\Lambda}/g_{N^*(1535)p\eta} =1.3\pm 0.3$. With
previous known value of $g_{N^*(1535)p\eta}$, the obtained new
value of $g_{N^*(1535)K\Lambda}$ is shown to reproduce recent
$pp\to pK^+\Lambda$ near-threshold cross section data as well.
Taking into account this large $N^*K\Lambda$ coupling in the
coupled channel Breit-Wigner formula for the $N^*(1535)$, its
Breit-Wigner mass is found to be around 1400 MeV, much smaller
than previous value of about 1535 MeV obtained without including
its coupling to $K\Lambda$. The implication on the nature of
$N^*(1535)$ is discussed.
\end{abstract}

\pacs{14.20.Gk, 13.30.Eg, 13.75.Jz}

\maketitle

The properties and the nature of the lowest spin-1/2 negative
parity ($J^P=1/2^-$) nucleon resonance $N^*(1535)$ are of great
interests in many aspects of light hadron physics. In conventional
constituent quark models, the lowest $1/2^-$ $N^*$ resonance
should be the first $L=1$ orbital excitation state. But it has
been a long-standing problem for these conventional constituent
quark models to explain why the mass of $N^*(1535)$ has a mass
higher than the lowest $J^P=1/2^+$ radial excitation state
$N^*(1440)$ \cite{Capstick1}. This was used to argue in favor of
the Goldstone-boson exchange quark models \cite{Glozman}. In the
recent Jaffe-Wilczek diquark picture \cite{Jaffe} for the $\theta$
pentaquark, a $J^P=1/2^-$ $N^*$ pentaquark of mass around 1460 MeV
is expected \cite{Zhu}. Another outstanding property of the
$N^*(1535)$ is its extraordinary strong coupling to $\eta N$
\cite{PDG}, which lead to a suggestion that it is a quasi-bound
($K\Sigma$-$K\Lambda$)-state \cite{Weise}. This picture predicts
also large effective couplings of $N^*(1535)$ to $K\Lambda$ and
$K\Sigma$ \cite{Oset}. Experiment knowledge on these kaon-hyperon
couplings is poor, partly because lack of data on experimental
side and partly due to the complication of various interfering
t-channel exchange contributions \cite{Mosel}. Better knowledge on
these couplings is definitely useful for understanding the nature
of $N^*(1535)$, the underneath quark dynamics, and also the
strangeness production in relativistic heavy-ion collisions as a
signature of the quark-gluon plasma \cite{Capstick2,Wu,Lee}.

Recently various $N^*$ production processes from $J/\psi$ decays
have been investigated by BES collaboration
\cite{Lihb,ppeta,Zoubs,Yanghx}. In the $J/\psi\to\bar pp\eta$
\cite{Lihb,ppeta} and $\psi\to\bar pK^+\Lambda + c.c.$
\cite{Zoubs,Yanghx} reactions, there are clear peak structures
with $J^P=1/2^-$ in the $p\eta$ and $K\Lambda$ invariant mass
spectra around $p\eta$ and $K\Lambda$ thresholds. A nature source
for the peak structures is $N^*(1535)$ coupling to $N\eta$ and
$K\Lambda$. In this letter, assuming the $1/2^-$ $K\Lambda$
threshold peak to be dominantly from the tail of the $N^*(1535)$,
we deduce the ratio between effective coupling constants of
$N^*(1535)$ to $K\Lambda$ and $p\eta$, $R\equiv
g_{N^*(1535)K\Lambda}/g_{N^*(1535)p\eta}$ from the new branching
ratio results from BES experiment on $J/\psi\to\bar pp\eta$ and
$\psi\to\bar pK^+\Lambda$, then check the compatibility with
recent $pp\to pK^+\Lambda$ near-threshold data \cite{cosy1,cosy2}.
Taking into account the large $N^*K\Lambda$ coupling in the
coupled channel Breit-Wigner formula for the $N^*(1535)$, we show
it gives a very large influence to the Breit-Wigner mass of the
$N^*(1535)$.

\begin{figure}[htbp] \vspace{-0.6cm}
\begin{center}
\includegraphics[scale=0.7]{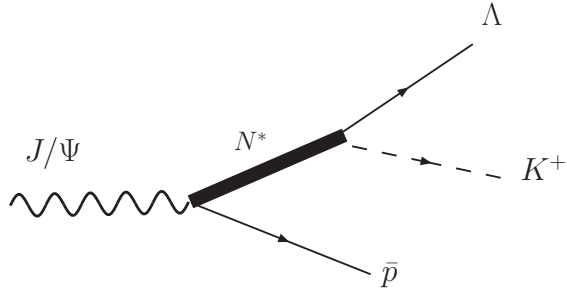}
 \caption{ Feynman diagram for $\psi\to\bar pK^+\Lambda$ through $N^*$ resonance}
 \label{psi}
\end{center}
\end{figure}

In the effective Lagrangian approach for the resonance isobar
model, the Feynman diagram for $\psi\to\bar pK^+\Lambda$ through
$N^*(1535)$ intermediate is shown in Fig.\ref{psi}. For
$\psi\to\bar pp\eta$, besides a similar diagram through
$N^*(1535)$, a diagram through $\bar N^*(1535)$ should be added
simultaneously. The relevant interaction Lagrangians are
\cite{Benmerrouche,Olsson}
\begin{eqnarray}
{\cal L}_{N^*\Lambda K} & = & -ig_{ N^*\Lambda K}
{\bar{\Psi}}_{\Lambda} \Phi_K \Psi_{N^*} + h.c. ,\\
{\cal L}_{N^*N \eta} & = & -ig_{N^* N \eta}
{\bar{\Psi}}_N  \Phi_\eta \Psi _{N^*}+h.c. ,\\
{\cal L}_{\psi  N N^* }^{(1)} & = & \frac {ig_{T}}{M_{N^*}+M_p}
{\bar{\Psi}}_{N^*} \gamma_5 \sigma_{\mu \nu} p^\nu_\psi \Psi _N \varepsilon^\mu+h.c. ,\\
{\cal L}_{\psi  N N^* }^{(2)} & = & -g_V {\bar{\Psi}}_{N^*}
\gamma_5 \gamma_\mu \Psi _N \varepsilon^\mu+h.c. \label{ignore}
\end{eqnarray}
where $\Psi_{N^*}$ represents the resonance $N^*(1535)$ with mass
$M_{N^*}$, $\Psi_N$ for proton with mass $M_p$ and
$\varepsilon^\mu$ for J/$\psi$ with four-momentum $p_\psi$.
According to \cite{ppeta}, the ${\cal L}_{\psi  NN^* }^{(2)}$ term
given by Eq.(\ref{ignore}) makes insignificant contribution for
$N^*(1535)$, hence we drop this kind of coupling in our
calculation. The amplitudes for $J/\psi\to\bar p K^+ \Lambda$ and
$\bar pp\eta$ via $N^*(1535)$ resonance are then
\begin{eqnarray}
M_{\psi\to\bar pK^+\Lambda} &=& \frac{i g_T g_{ N^*K\Lambda}}
{M_{N^*}+M_p} \bar u (p_\Lambda,s_\Lambda)(\sla{p}_{N^*}+m_{N^*})
BW(p_{N^*}) \gamma_5 \sigma_{\mu \nu} p^\nu_\psi
\varepsilon^\mu v(p_{\bar p},s_{\bar p}) , \\
M_{\psi\to\bar pp\eta} &=& \frac{i g_T g_{ N^*N\eta}}
{M_{N^*}+M_p} \bar u (p_p,s_p) [(\sla{p}_{N^*}+m_{N^*})
BW(p_{N^*})
\gamma_5 \sigma_{\mu \nu} p^\nu_\psi \varepsilon^\mu + \nonumber\\
& & \quad\quad\quad\quad\quad\quad \gamma_5 \sigma_{\mu \nu}
p^\nu_\psi \varepsilon^\mu (-\sla{p}_{\bar N^*}+m_{N^*})
BW(p_{\bar N^*}) ]v(p_{\bar p},s_{\bar p}),
\end{eqnarray}
respectively. Here $BW(p_{N^*})$ is the  Breit-Wigner formula for
the $N^*(1535)$ resonance
\begin{equation}
BW(p_{N^*})={1\over M^2_{N^*} - s - i M_{N^*}\Gamma_{N^*}(s)}
\end{equation}
with $s=p^2_{N^*}$. According to PDG \cite{PDG}, the dominant
decay channels for the $N^*(1535)$ are $N\pi$ and $N\eta$. For a
resonance with mass close to some threshold of its dominant decay
channel, the approximation of a constant width is not very good.
Since the $N^*(1535)$ is quite close to the $\eta N$ threshold, we
take the commonly used phase space dependent width for the
resonance as the following
\begin{equation}
\Gamma_{N^*}(s)=\Gamma^0_{N^*} \left( 0.5 \frac {\rho_{\pi
N}(s)}{\rho_{\pi N}(M^2_{N^*})}+0.5 \frac {\rho_{\eta
N}(s)}{\rho_{\eta N}(M^2_{N^*})}\right)=\Gamma^0_{N^*} \left[ 0.8
\rho_{\pi N}(s)+2.1\rho_{\eta N}(s)\right],\label{gammar}
\end{equation}
where $\rho_{\pi N}(s)$ and $\rho_{\eta N}(s)$ are the phase space
factors for $\pi N$ and $\eta N$ final states, respectively, e.g.,
\begin{equation}
\rho_{\eta N}(s) = \frac{2q_{\eta N}(s)}{\sqrt{s}} =
\frac{\sqrt{(s-(M_N+M_\eta)^2)(s-(M_N-M_\eta)^2)}}{s}
\end{equation}
where $q_{\eta N}$ is the momentum of $\eta$ or N in the
center-of-mass system of $\eta N$. According to PDG \cite{PDG},
$M_{N^*}\approx 1535 MeV$ and
$\Gamma^0_{N^*}=\Gamma_{N^*}(M^2_{N^*})\approx 150 MeV$.

From the amplitudes given above, we can calculate the decay widths
of $J/\psi\to\bar pK^+\Lambda$ and $J/\psi\to\bar pp\eta$ via
$N^*(1535)$ resonance, and get their ratio as
\begin{equation}
\frac { \Gamma(\psi\to\bar pN^*\to\bar pK^+\Lambda)} { \Gamma(\psi
\to\bar pN^* + p\bar N^*\to\bar pp\eta)}=\frac{1}{12.6}\left
|\frac{g_{N^*K\Lambda}}{g_{N^*N\eta}}\right |^2 .\label{frac1}
\end{equation}
On the other hand, from PDG and recent BES results, we have
$J/\psi$ decay branching ratio for the $\bar pK^+\Lambda$ channel
as $(0.89\pm 0.16)\times 10^{-3}$ \cite{PDG} with $(15\sim 22)\%$
\cite{Yanghx} via the near threshold $N^*$ resonance and for the
$\bar pp\eta$ channel as $(2.09\pm 0.18)\times 10^{-3}$ \cite{PDG}
with $(56\pm 15)\%$ \cite{ppeta} via the $N^*(1535)$ resonance.
Therefore
\begin{equation}
\frac { \Gamma(\psi\to\bar pN^*\to\bar pK^+\Lambda)} { \Gamma(\psi
\to\bar pN^* + p\bar N^*\to\bar pp\eta)}=\frac{(0.89\pm
0.16)\times (15\sim 22)}{(2.09\pm 0.18)\times (56\pm 15)}.
\label{frac2}
\end{equation}
From Eq(\ref{frac1}) and Eq(\ref{frac2}), we get
\begin{equation}
R\equiv\left | \frac{g_{N^*(1535)K\Lambda}}{g_{N^*(1535)N \eta}}
\right | \approx 1.3 \pm 0.3. \label{relation}
\end{equation}
Previous knowledge on this ratio from $\pi N\to K\Lambda$ and
$\gamma N\to K\Lambda$ reactions is poor. While Ref.\cite{Mosel}
gave a range of $0.8\sim 2.6$, others found the contribution from
the $N^*(1535)$ is not important for reproducing the data
\cite{Lee}. It seems that those data are not sensitive to the
$N^*(1535)$ contribution due to the complication of various
interfering t-channel contributions which are absent in the
$J/\psi$ decays. Another relevant reaction is $pp\to pK^+\Lambda$.
Some very precise near-threshold data are now available from COSY
experiments \cite{cosy1,cosy2}. In the following we will check the
compatibility of the large R value given by Eq.(\ref{relation})
with the recent $pp\to pK^+\Lambda$ near-threshold data.

\begin{figure}[htbp] \vspace{-0.cm}
\begin{center}
\includegraphics[scale=0.7]{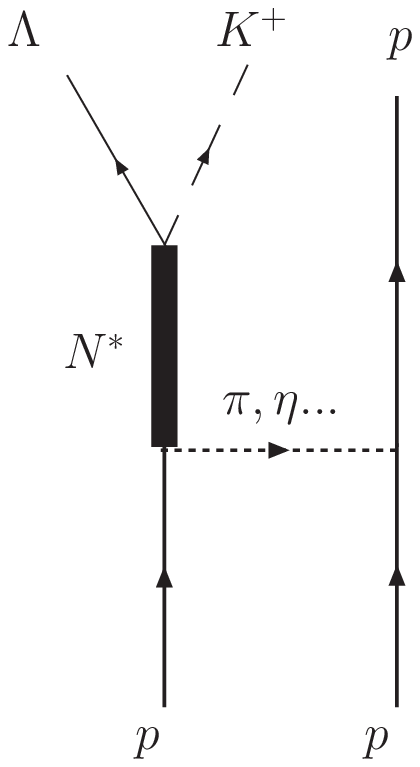}
\includegraphics[scale=0.7]{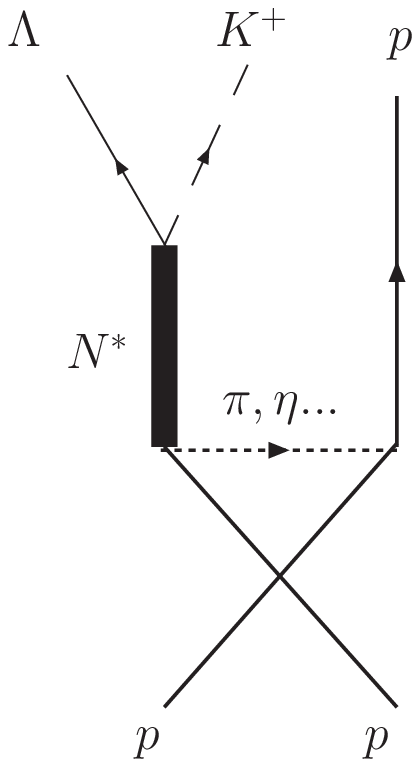}
 \caption{ Feynman diagrams for reaction $pp\to pK^+\Lambda$ }
 \label{block1}
\end{center}
\end{figure}

The relevant Feynman diagrams for the process $pp\to pK^+\Lambda$
are shown in Fig.\ref{block1}.  Since we are mainly interested in
the near-threshold behavior where contribution from $\pi$ and
$\eta$ meson exchange dominates \cite{Tsushima}, here for
simplicity we ignore the small contribution from heavier mesons.
We adopt the relevant effective Lagrangian and form factors used
in Ref. \cite{Tsushima}.

\begin{figure}[htbp] \vspace{-0.cm}
\begin{center}
\includegraphics[scale=0.75]{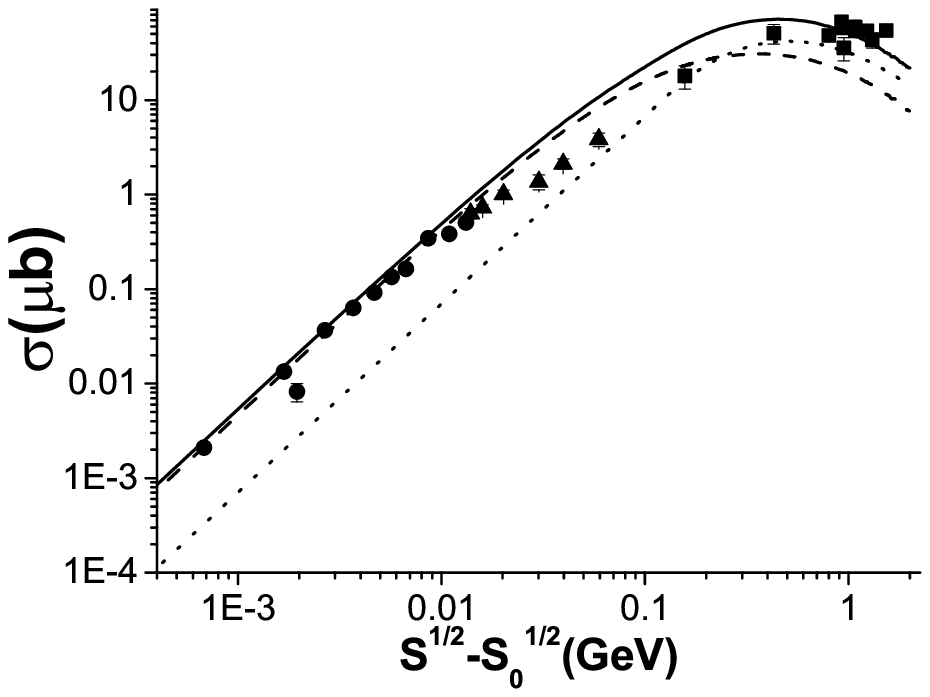}
\includegraphics[scale=0.75]{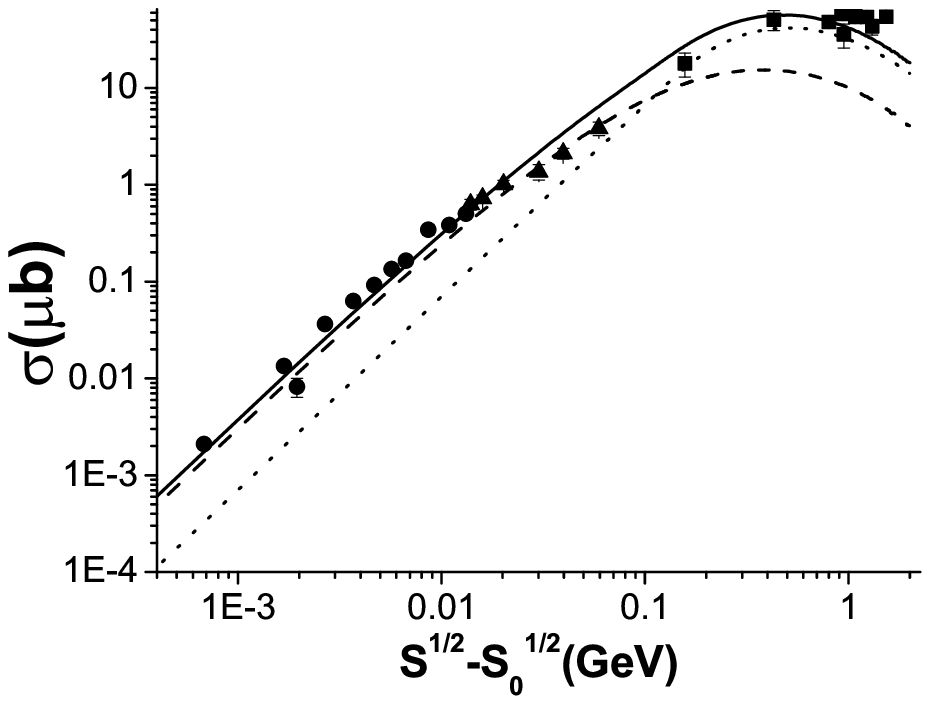}%
\caption{ The cross section of the reaction $pp\to pK^+\Lambda$ as
a function of the excess energy with data from Refs.\cite{cosy1}
(circle), \cite{cosy2} (triangle) and \cite{Landolt} (square). The
dashed and dotted lines represent the contribution from
$N^*(1535)$ and other $N^*$ resonances, respectively. The solid
line is the sum. The left and right graphs are the results without
and with including $\Lambda K$ term in the $\Gamma_{N^*}(s)$ for
$N^*(1535)$. }
 \label{result}
\end{center}
\end{figure}

First we have reproduced the results of Ref.\cite{Tsushima} by
including $N^*(1650)1/2^-$, $N^*(1710)1/2^+$ and $N^*(1720)3/2^+$
resonances. Their prediction prior COSY data \cite{cosy1,cosy2} is
shown by the dotted line in Fig.\ref{result}, which is obviously
underestimating the near-threshold data of COSY. In their work,
all parameters have been fixed by previous study on other relevant
reactions. A natural reason for the underestimation is their
ignorance of the contribution from $N^*(1535)$.  Here we calculate
the contribution from $N^*(1535)1/2^-$ for the process. The
coupling constants for the vertices $N^*(1535)N\pi$ and
$N^*(1535)N\eta$ are determined by the relevant partial decay
width \cite{PDG}. Then the coupling constant for the $N^*K\Lambda$
is obtained by our new result $\left |
g_{N^*(1535)K\Lambda}/g_{N^*(1535)N \eta} \right | = 1.3$ from BES
data.  The result is shown by the dashed line in Fig.\ref{result}
(left). Adding the contribution to the previous results of
Ref.\cite{Tsushima}, the solid line in Fig.\ref{result} (left)
reproduces the COSY near-threshold data very well. So the ratio
given by Eq.(\ref{relation}) is also compatible with the data on
$pp\to pK^+\Lambda$. Note we have not introduce any free
parameters in this calculation.

The large $\left | g_{N^*(1535)K\Lambda}/g_{N^*(1535)N \eta}
\right |$ ratio has important implications on other properties of
the $N^*(1535)$. First, in previous calculations, the coupling of
$N^*(1535)$ to $K\Lambda$ channel is usually ignored in the
Breit-Wigner formula for the $N^*(1535)$. Considering this
coupling, the width in its Breit-Wigner formula should be
\begin{equation}
\Gamma_{N^*}(s)=\Gamma^0_{N^*} \left[ 0.8 \rho_{\pi
N}(s)+2.1\rho_{\eta N}(s)+3.5\rho_{\Lambda K}(s)
\right]\label{gammar2}
\end{equation}
instead of Eq.(\ref{gammar}). In order to give a similar
Breit-Wigner amplitude squared $|BW(p_{N^*})|^2$ as using
Eq.(\ref{gammar}) with $M_{N^*}=1535 MeV$ and $\Gamma^0_{N^*}= 150
MeV$, we need $M_{N^*}\approx 1400 MeV$ and $\Gamma^0_{N^*}= 270
MeV$ when using Eq.(\ref{gammar2}). Note that the two-body phase
space factors $\rho_{\eta N}(s)$ and $\rho_{\Lambda K}(s)$ are
extended to below their corresponding thresholds to be pure
imaginary as the Flatt\'e formulation for $f_0(980)$ meson
\cite{Flatte}.

\begin{figure}[htbp] \vspace{-0.cm}
\begin{center}
\includegraphics[scale=0.5,trim=0 150 0 0]{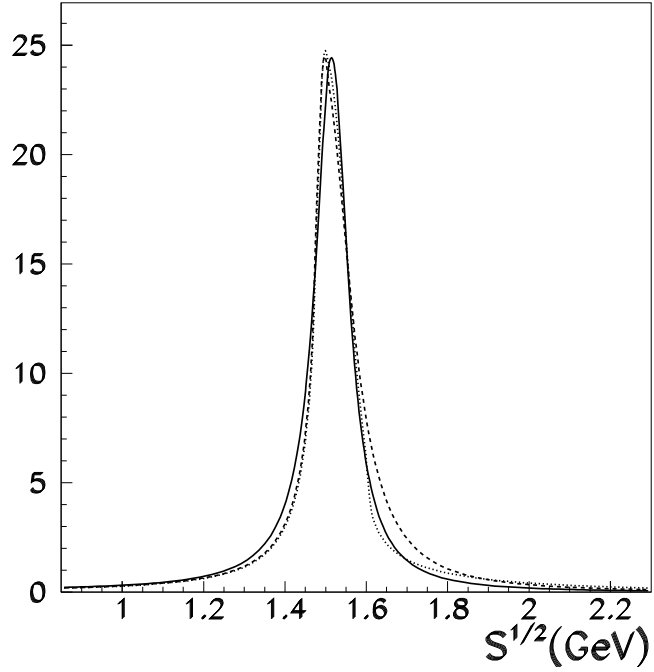}
 \caption{Breit-wigner amplitude squared vs $s^{1/2}$ with a constant
width (solid line), energy-dependent width without (dashed line)
and with (dotted line) $\Lambda K$ channel contribution  included.
}
 \label{bw}
\end{center}
\end{figure}

In Fig.\ref{bw}, we show the Breit-wigner amplitude squared vs
$s^{1/2}$ for the two cases without (dashed line) and with (dotted
line) $\Lambda K$ channel contribution included in the
energy-dependent width for the $N^*(1535)$. As a comparison, we
also show the case assuming a constant width $\Gamma_{N^*}(s)= 98
MeV$ with $M_{N^*}=1515 MeV$ (solid line). The three kinds of
parametrization for the $N^*(1535)$ amplitude give a similar
amplitude squared,  hence do not influence much on previous
calculations on various processes involving the $N^*(1535)$
resonance by using the Breit-Wigner formula without including the
$\Lambda K$ channel in the width. As an example, we show in
Fig.\ref{result} (right) the results including the $\Lambda K$
channel in $\Gamma_{N^*}(s)$. Comparing results in
Fig.\ref{result} (left) without including the $\Lambda K$ channel
in $\Gamma_{N^*}(s)$, while the fit to the data for the energies
between 10 MeV and $400 MeV$ improves a little bit, the over all
shape looks very similar. However, the important point is that by
including the large $N^*K\Lambda$ coupling in the coupled channel
Breit-Wigner formula for the $N^*(1535)$, its Breit-Wigner mass is
reduced to be around 1400 MeV, much smaller than previous value of
about 1535 MeV obtained without including its coupling to
$K\Lambda$. This will have important implication on various model
calculations on its mass.

The second important implication of the large $N^*K\Lambda$
coupling is that the $N^*(1535)$ should have large $s\bar s$
component in its wave function. It has been suggested to be a
quasi-bound ($K\Sigma$-$K\Lambda$)-state \cite{Weise}. Based on
this picture, the effective coupling of $N^*(1535)$ to $K\Lambda$
is predicted to be about $0.5\sim 0.7$ times of that for
$N^*(1535)$ to $\eta N$ \cite{Oset}, which is about a factor 2
smaller than the value obtained here. Alternatively, the
strangeness may mix into the $N^*(1535)$ in the form of some
pentaquark configuration \cite{Riska}. According to
Ref.\cite{Riska}, the $[4]_X[31]_{FS}[211]_F[22]_S (qqqs\bar s)$
pentaquark configuration has the largest negative flavor-spin
dependent hyperfine interaction for $1/2^-$ $N^*$ resonance. Hence
the $1/2^-$ $N^*(1535)$ resonance may have much larger $(qqqs\bar
s)$ pentaquark configuration than $1/2^+$ $N^*$ resonances, for
which the penta-quark configurations with the largest negative
flavor-spin dependent hyperfine interaction are non-strange ones,
such as $[31]_X[4]_{FS}[22]_F[22]_S (qqqq\bar q)$ configuration
\cite{Riska}. This will result in a large $N^*K\Lambda$ coupling.
A concrete calculation in this picture should be very useful for
understanding the nature of the $N^*(1535)$. A recent study of the
strangeness in the proton \cite{Zou} suggests that the strangeness
in the nucleon and its excited states $N^*$ are most likely in the
form of pentaquark instead of meson-cloud configurations.

Another implication of the large $N^*(1535)K\Lambda$ coupling is
that many previous calculations on various $K\Lambda$ production
processes without including this coupling properly should be
re-examined. A proper treatment of the $N^*(1535)$ contribution
may help to extract properties of other $N^*$ resonances more
reliably.

In summary, from the recent BES data on $J/\psi\to\bar pp\eta$ and
$\psi\to\bar pK^+\Lambda$, the
$g_{N^*(1535)K\Lambda}/g_{N^*(1535)p\eta}$ ratio is deduced to be
$1.3\pm 0.3$ which is also compatible with data from $pp\to
pK^+\Lambda$, $\pi p\to K\Lambda$ and $\gamma p\to K\Lambda$
processes. By including the large $N^*(1535)K\Lambda$ coupling
into the Breit-Wigner formula for the $N^*(1535)$, a much lower
Breit-Wigner mass ($\sim 1400 MeV$) is obtained for the
$N^*(1535)$.  These new properties have important implication on
the nature of the lowest negative-parity $N^*$ resonance. The
$N^*(1535) 1/2^-$ could be the lowest $L=1$ orbital excited $(3q)$
state with a large admixture  of $[4]_X[31]_{FS}[211]_F[22]_S
(qqqs\bar s)$ pentaquark component while the $N^*(1440)$ could be
the lowest radial excited $(3q)$ state with a large admixture of
$[31]_X[4]_{FS}[22]_F[22]_S (qqqq\bar q)$ pentaquark component.
While the lowest $L=1$ orbital excited $(3q)$ state should have a
mass lower than the lowest radial excited $(3q)$ state, the
$(qqqs\bar s)$ pentaquark component has a higher mass than
$(qqqq\bar q)$ pentaquark component. This makes the $N^*(1535)$
having an almost degenerate mass with the $N^*(1440)$.

\bigskip
We thank S.Jin, D.Riska and A.Sibirtsev for useful discussion.
This work is partly supported by the National Nature Science
Foundation of China under grants Nos. 10225525, 10435080 and by
the Chinese Academy of Sciences under project No. KJCX2-SW-N02.

\end{document}